# Gusts and Shear Within Hurricane Eyewalls Can Exceed Offshore Wind-Turbine Design Standards

**Rochelle P. Worsnop[1], Julie K. Lundquist[1,2], George H. Bryan[3], Rick Damiani[2], and Walt Musial[2]**

[1]Department of Atmospheric and Oceanic Sciences, University of Colorado at Boulder, Boulder, Colorado, USA.

[2]National Renewable Energy Laboratory, Golden, Colorado, USA.

[3]National Center for Atmospheric Research, Boulder, Colorado, USA.

Corresponding author: Rochelle Worsnop (rochelle.worsnop@colorado.edu)

**Key Points:**

- Large-eddy simulations quantify hurricane gusts, gust factors, and wind direction shifts

- Category 3 or greater hurricane mean wind speed, gusts, and gust factor in eyewall exceed current design thresholds

- Inclusion of veer in turbine load calculations is recommended for hurricane-prone regions

**Abstract**

Offshore wind energy development is underway in the U.S., with proposed sites located in hurricane-prone regions. Turbine design criteria outlined by the International Electrotechnical Commission do not encompass the extreme wind speeds and directional shifts of hurricanes stronger than a Category 2. We examine the most turbulent portion of a hurricane (the eyewall) using large-eddy simulations with Cloud Model 1 (CM1). Gusts and mean wind speeds near the eyewall exceed the current design threshold of 50 m s$^{-1}$ mean wind and 70 m s$^{-1}$ gusts for Class I turbines. Gust factors are greatest at the eye-eyewall interface. Further, shifts in wind direction at wind turbine hub height suggest turbines must rotate into the wind faster than current practice. Although current design standards omit mention of wind direction change across the rotor layer, large values (15-50 deg) suggest that veer should be considered in design standards.

**1 Introduction**

Offshore wind energy generation in the U.S. began with a 20-m tall, 20 kW test-turbine deployed ~ 4 km off Maine's coast [*Russo,* 2014]. The first utility-scale wind farm, with generation capacity of 30 MW, has been completed off the coast of Block Island, Rhode Island [*Justin Gillis*, 2016]. Some offshore wind farm sites are proposed in regions with hurricane return periods less than the expected lifetime of a wind farm (20 years) [*Keim et al.*, 2007; *Russo*, 2014]. According to the National Hurricane Center (NHC), a return period of a major hurricane (1-min sustained winds ≥ 49 m s$^{-1}$ at 10 m AGL) is as short as 16 years along the North Carolina coast [*NHC*, 2016], where wind turbine development is likely. Three-second gusts exceeding this wind speed threshold can occur even more frequently at a given location [*Neumann*, 1991]. Hurricane winds pose a substantial risk to potential offshore wind turbines deployed in hurricane-prone regions as demonstrated by the destruction of turbines during Typhoons Maemi (2003) and Usagi (2013) [*Chen and Xu*, 2016]. Therefore, hurricane-tolerant turbine designs are now being considered to address this risk [*DOE*, 2015].

Current design standards for offshore wind turbines do not account for extreme winds associated with tropical cyclones. We demonstrate that mean wind speed, 3-sec gusts, gust factor, and shifts in wind direction exceed the current design criteria for the strongest class of wind turbines (Class I) outlined in the International Electrotechnical Commission (IEC) 61400-1 [2007] standard, suggesting that modifications are needed to account for even harsher environments. Because the classes of turbines in the IEC standard are not intended to cover tropical storm environments [*IEC*, 2007], a special design class is therefore needed for turbines in hurricane-prone regions. The American Petroleum Institute also calls for the understanding of the extreme wind conditions of hurricanes, including mean wind profiles, gusts, gust factors, turbulence spectra and coherence [*API*, 2010].

Important gaps undermine the research community's understanding of the hurricane boundary layer (HBL). Mean wind profiles of the horizontal wind speed follow a logarithmic wind profile from $\sim 20$ m $- 300$ m (ASL) [*Powell et al.*, 2003; *Vickery et al.*, 2009], consistent with the IEC's standard logarithmic profile. *Vickery and Skerlj* [2005] found that the IEC's gust factor is generally a good representation of hurricane gust factors. However, their calculations apply mostly at 10 m ASL, for wind speeds 60 m s$^{-1}$ or weaker. They did not examine gust factors at hub height, which are needed to determine the loads turbines experience in major hurricanes. *Worsnop et al.* [2016] assessed turbulence spectra and spatial coherence across a theoretical wind turbine within a simulated HBL. Their results differ from the spectral and coherence curves in the IEC standard, suggesting that unique characteristics of wind speed and direction within the HBL may also violate current design standards.

To capture the most extreme winds, data are required from within the eyewall, the turbulent region surrounding the eye of the hurricane [*Powell and Cocke*, 2012]. However, offshore hurricane wind data at turbine heights (below 200 m ASL) are extremely limited. Reconnaissance flights are normally flown at 1.5 - 3 km ASL [*French et al.*, 2007; *Cione et al.*, 2016], offshore towers are sparse [*Archer et al.*, 2013] and usually incur damage from direct hits, and dropsonde data are spatially limited within the storm [*Stern et al.*, 2016]. However, large-eddy simulations (LES) can provide simulated winds within the eyewall of the hurricane at turbine heights with high spatial ($\sim$30 m) and temporal resolution ($\sim$0.18 s). With continuous data across the model domain, we can also examine the radial dependence of wind speed and direction to determine the most problematic regions within the hurricane for wind turbines.

Herein, we use LES output to provide critical data to revise offshore wind turbine design standards. In section 2, we discuss the model configuration and data aggregation methods. In section 3, we quantify expected gusts and gust factors at hurricane radii smaller than 30 km. We discuss shifts in wind direction at hub height in section 4 and veer across the rotor layer in section 5. Lastly, we summarize our findings and offer suggestions to modify the current design standard in section 6.

## 2 LES of a hurricane

We simulate an idealized Category 5 hurricane, a worst-case scenario for wind turbines: damage increases exponentially with wind speed [*Landsea*, 1993]. We use Cloud Model 1 (CM1) [*Bryan and Rotunno*, 2009; *Bryan et al.*, 2016]. This idealized simulation is not intended to model a specific hurricane, but is based on a Category 5 hurricane, Felix (2007). The simulation's outer domain (3000 x 3000 x 25 km) encompasses the entire hurricane (eye, eyewall, and outer bands). Nested within this outer domain, a fine-mesh domain (80 km x 80km x 3km) with horizontal (vertical) grid spacing of 31.25 m (15.625 m) resolves the turbulent winds within the inner core, including the eye and eyewall. We output data every 0.1875 s time step at mock towers located every kilometer in x and y (Figure 1) and at every altitude, 7.81 m to 507.81 m ASL.

This CM1 simulation is initialized using output from an axisymmetric model, plus random perturbations, as in *Worsnop et al.* [2016] and *Richter et al.* [2016]. The simulation reaches statistical steady-state after four hours. We analyze wind fields from an additional ten minutes, which is the averaging period commonly used in the IEC standard and recommended by the World Meteorological Organization (WMO) [*Harper et al.*, 2010]. LES are inherently computationally expensive; this simulation required more than 500,000 core hours (one week of wall-clock time using 4,096 cores). Validation of CM1 compared to hurricane observations is discussed in *Worsnop et al.* [2016]. The model configuration is identical to the "Complex" simulation of *Worsnop et al.* [2016]; here, we double the temporal and spatial resolution.

We calculate 10-min mean wind speeds, 3-sec gusts, gust factor, directional shifts at hub height (~ 100 m), and veer at each mock tower location shown in Figure 1. We then aggregate the towers into 1-km-radii bins to obtain a representative sample at each radii. Finally, we take the maximum value of these variables at each radius to assess the strongest winds conditions a wind turbine would experience in a major hurricane.

### 3 Hurricane gusts and gust factor

Flow in the HBL is not homogeneous: greatest wind speeds are found within the eyewall. The peak 3-sec gust quantifies the highest 3-sec average wind speed observed within a longer interval, here 10 min [*Harper et al.*, 2010], and is used to estimate loads that turbines would endure. Gusts exceeding 70 m s$^{-1}$ at altitudes across the rotor layer are problematic [*IEC,* 2007] and may cause significant damage; we find these gusts occur within the eyewall and just outside of the eyewall of major hurricanes (Figure 2b-d), with some gusts exceeding 100 m s$^{-1}$. For this storm, this critical region spans ~10 km ($R$ = 10-20 km). These wind speeds agree with maximum observed values in an analysis of ~12,000 dropsondes from tropical cyclones by [*Stern et al.*, 2016]; they found that extreme horizontal wind speeds ($\geq$ 90 m s$^{-1}$) and updrafts ($\geq$ 10 m s$^{-1}$) can occur within the eyewall at altitudes as low as 100 m ASL. Outside the eyewall region, gusts fall below 70 m s$^{-1}$.

By considering the maximum 10-min mean wind speed and 3-sec gusts at each radius (Figure 3a), we identify regions where the winds exceed current Class I design thresholds (i.e., 50 m s$^{-1}$ mean wind and 70 m s$^{-1}$ peak gusts) and thus where wind turbines would fail if designed to current standards . Even these thresholds may be too lenient: wind turbines along Japan's coast were severely damaged in mean wind speeds below this threshold (~38 m s$^{-1}$): the peak gust reached 74 m s$^{-1}$ during Typhoon Maemi in 2003 [*Ishihara et al.*, 2005]. As seen in the peak gust contours in Figure 3 (black lines), gusts can exceed 100 m s$^{-1}$ over a range of ~17 km, suggesting that turbines developed to withstand loads induced by major hurricanes should anticipate gusts much higher than the current gust threshold (or that turbine destruction should be assumed for events like this one).

Gust factor, $G_{t,T_0}$, estimates the expected peak gust when only mean wind speed, $V_{T_0}$, is known:

$$G_{t,T_0} = \frac{V_{\tau,T_0}}{V_{T_0}}, \tag{1}$$

where $V_{\tau,T_0}$ is the highest 3-sec mean (gust) that occurs within 10 min ($\tau = 3$ sec, $T_0 = 600$ sec) [*Harper et al.*, 2010]. If the 3-sec gust value at hub height is desired for a hurricane with a certain mean wind speed, then it is imperative to have a gust factor representative of wind speeds at hub height. *Vickery and Skerlj* [2005] determined hurricane gust factors from onshore and offshore observations collected below 40 m and only for wind speeds less than 60 m s$^{-1}$. Here, we examine hurricane gust factors at turbine heights and at different radii to show where the highest gust factors can be expected (Figure 3b).

The highest gust factors (~ 1.7) outside of the quiescent eye occur at the eye-eyewall interface (here, $R$ = 9-11 km) just inward of the peak gusts in the hurricane (we ignore gust factors within the eye because mean wind speeds there are too weak to impact wind turbines). Additionally, gust factors of ~1.4 or greater occur below 30 m ASL, on average, at radii just outside of the eyewall (here, $R$ > 11 km). Generally, the gust factor is ≤ 1.4 (the standard IEC value corresponding to a mean wind speed of 50 m s$^{-1}$ and a gust of 70 m s$^{-1}$) outside of the eyewall (except below ~30 m ASL). This result is consistent with *Vickery and Skerlj* [2005] who showed that gust factors over the open water were generally less than 1.4 for wind speeds less than 60 m s$^{-1}$. However, gust factors as high as 1.7, combined with high wind speeds occur in the eyewall, and values of 1.5 could be seen close to the surface even outside the eyewall. Our data indicate that a gust factor of 1.4 is adequate for turbine design in most regions of the hurricane, but underestimates the eyewall region and regions outside of the eyewall below ~30 m ASL.

## 4 Directional wind shifts at hub height and yaw misalignment

Wind direction can shift 180 deg during a hurricane passage within 0.5 to 1.5 hours [*Clausen et al.*, 2007]. While turbines can yaw, or rotate into the mean wind direction, to prevent harsh loads on the turbine, abrupt changes in wind direction may affect turbine survival. Edgewise vibrations induced by yaw misalignment [*Fadaeinedjad et al.*, 2009] can damage turbine blades and induce buckling of the tower. Yaw misalignment caused turbines to fail at a wind farm in China during the passage of Typhoon Dujuan (2003), even when wind speeds were below the design speed [*Clausen et al.*, 2007].

We find large shifts in wind direction at hub height (Figure 4). The largest shifts occur within the eyewall (here, ~10 km) perhaps due to coherent vortices ("mesovortices") between the eye and eyewall [*Aberson et al.*, 2006]. Turbines typically yaw based on the recorded average (usually 10 min) change in wind direction and on the current wind speed. However, for higher wind speeds, a shorter averaging time and shorter yaw response-time can be used. The tails of the distributions in Figure 4 reveal that the wind direction can shift 10-30 deg in durations less than 10 min even outside of the eyewall. Yaw misalignment could occur frequently in these storms if the yaw system is not designed to sense directional shifts at one minute or less. Abrupt changes in wind direction at hub height near the HBL eyewall suggest that a faster yaw response than 10 min may be needed.

## 5 Wind veer across the rotor layer

Current design standards do not address veer, the change in wind direction across the vertical rotor layer, even though veer may affect loads on turbines and has been shown to affect power production [*Walter et al.*, 2009; *Vanderwende and Lundquist*, 2012]. Varying wind direction across different portions of the turbine can cause additional stress to the turbine, which can lead to mechanical failure. We calculate the maximum average veer relative to hub height (100 m) over averaging periods from three sec to one min (Figure 5). Within 200 m ASL, we find that the wind direction can change > 35 deg (maximum of 55 deg) with respect to wind direction at hub height for periods ≤ 10 sec (Figure 5a, b). Veer ranges from 5-15 deg, particularly below 50 m for averaging periods of 30 sec and 1 min (Figure 5c, d). This strong veer demonstrates that wind turbines will endure swift changes in wind direction, across the vertical rotor layer, on the order of 1 min or less during a hurricane eyewall passage. Testing the influence of an average veer of roughly 15 deg in load simulators such as FAST [*Jonkman and Buhl Jr.*, 2005] would reveal how veer impacts loading on the turbine and whether manufacturers should include veer as a more vital component in the design process for offshore wind turbines.

## 6 Conclusion

We examined gusts, gust factor, and wind direction changes in the hurricane boundary layer (HBL) and compared these values to those currently used in the IEC wind turbine design standard. We represented the HBL with large-eddy simulations of an idealized Category 5 hurricane using Cloud Model 1 (CM1). Results indicate that conditions outside of the design standards would be encountered by wind turbines experiencing the eyewall and just outside the eyewall regions of a Category 5 hurricane; turbines built to current design standards would incur structural damage.

Mean wind speed and 3-sec gusts are greatest in the turbulent eyewall of the hurricane. Within the eyewall and just outside of the eyewall, winds exceed the current turbine design thresholds of 50 m s$^{-1}$ mean wind and 70 m s$^{-1}$ peak gust. Mean wind speeds (gusts) can exceed 90 m s$^{-1}$ (100 m s$^{-1}$) within the eyewall, consistent with observations [*Stern et al.*, 2016], suggesting that either design standards or expected turbine lifetimes be modified to account for extreme conditions within a hurricane.

We also analyzed gust factors at multiple radii and at heights pertinent to wind turbines (< 200 m ASL). The largest gust factors occur at the interface between the eye and eyewall, just inward of the peak gusts. While the majority of the hurricane gust factors, away from the eyewall, are similar to a previous observational study [*Vickery and Skerlj* 2005], the gust factor ≥ 1.7 within the eyewall. While the eyewall makes up a small fraction of the total hurricane area, *Kimball and Mulekar* [2004] found, from a climatology of Atlantic tropical cyclones, that the median radius of maximum winds (RMW) is ~55 km, which could encompass most of a wind farm experiencing a direct strike. Additionally, for radii outside of the eyewall and RMW, the gust factor is greatest at altitudes less than 50 m ASL. At these locations, the gust factor can exceed 1.4, the value used in the IEC standard to convert a reference wind speed of 50 m s$^{-1}$ to a 3-sec gust. A value of 1.5 may be more accurate to estimate gusts at the lower sections (~ 30 m ASL) of wind turbines outside the eyewall of a Category 5 hurricane.

Wind direction shifts in the HBL can lead to significant yaw misalignment. In our simulations, wind directions can shift 10-30 deg over durations less than 10 min. Turbines should be able to respond to directional shifts on these shorter time-scales to avoid damaging loads. Finally, we quantified the absolute average veer for a typical turbine, and found shifts of 35 deg or greater from the hub to the tip of the rotor layer for periods of 3 and 10 sec. For averaging periods of 30 sec and 1 min, the veer is weaker, but can reach 15 deg. Veer across the turbine is not accounted for in the current design standard, but these results suggest that its influence should be tested in load simulators to determine if veer should be an essential component of turbine load estimations during hurricane passages.

These results are intended to guide turbine developers in the design of robust offshore wind turbines for hurricane-prone regions or in the quantification of financial risk for those offshore wind turbines. Investigation of the actual turbine loads induced by the gusts, veer, and possible yaw misalignments discussed within are needed to determine what modifications are required to build turbines to withstand major hurricanes. Incorporating these LES into turbine loads simulators as in *Sim et al.* [2012] and *Park et al.* [2015] and accounting for storm surge [*Jordan and Clayson*, 2008] and breaking waves [*Suzuki et al.*, 2014; *Hara and Sullivan*, 2015] near these offshore wind turbines would be a viable next step.

**Acknowledgments and Data**

This material is based upon work supported by the National Science Foundation under Grant No. DGE-1144083. We also acknowledge high-performance computing support from Yellowstone (ark:/85065/d7wd3xhc) provided by NCAR's Computational and Information Systems Laboratory and sponsored by the National Science Foundation. NREL is a national laboratory of the U. S. Department of Energy, Office of Energy Efficiency and Renewable Energy, operated by the Alliance for Sustainable Energy, LLC. Partial funding for this work was provided by the U. S. Department of Energy, Office of Energy Efficiency and Renewable Energy, Wind and Water Power Technologies Office. Outputs of the simulation used in this study can be obtained by contacting the first author.

**References**

Aberson, S. D., M. Black, M. T. Montgomery, and M. Bell (2006), Hurricane Isabel (2003): New Insights Into the Physics of Intense Storms. Part II: Extreme Localized Wind, *Bull. Am. Meteorol. Soc.*, *87*(10), 1349–1354, doi:10.1175/BAMS-87-10-1349.

American Petroleum Institute (2010), API BULLETIN 2INT-MET-Interim Guidance on Hurricane Conditions in the Gulf of Mexico Upstream Segment IEC (International Electrotechnical Commission). Offshore requirements for Wind Turibnes. IEC 61400-3 Ed1-Committee Draft 88/257/CD. Geneva, Switzerland:IEC. Available online at [https://law.resource.org/pub/us/cfr/ibr/002/api.2int-met.2007.pdf].

Archer, C. L. et al. (2013), Meteorology for Coastal/Offshore Wind Energy in the United States: Recommendations and Research Needs for the Next 10 Years, *Bull. Am. Meteorol. Soc.*, *95*(4), 515–519, doi:10.1175/BAMS-D-13-00108.1.

Bryan, G. H., and R. Rotunno (2009), The Maximum Intensity of Tropical Cyclones in Axisymmetric Numerical Model Simulations, *Mon. Weather Rev.*, *137*(6), 1770–1789, doi:10.1175/2008MWR2709.1.

Bryan, G. H., R. Worsnop, J. K. Lundquist, and J. A. Zhang (2016), A Simple Method for Simulating Wind Profiles in the Boundary Layer of Tropical Cyclones, *Bound.-Layer Meteor.* (in press). doi:10.1007/s10546-016-0207-0. Available online at [http://www2.mmm.ucar.edu/people/bryan/papers/blm_simple_tc_boundary_layer_v3.pdf].

Chen, X., and J. Z. Xu (2016), Structural failure analysis of wind turbines impacted by super typhoon Usagi, *Eng. Fail. Anal.*, *60*, 391–404, doi:10.1016/j.engfailanal.2015.11.028.

Cione, J. J., E. A. Kalina, E. W. Uhlhorn, A. M. Farber, and B. Damiano (2016), Coyote Unmanned Aircraft System Observations in Hurricane Edouard (2014), *Earth Space Sci.*, 2016EA000187, doi:10.1002/2016EA000187.

Clausen, N.-E., A. Candelaria, S. Gjerding, S. Hernando, P. Norgard, S. Ott, and N.-J. Tarp-Johansen (2007), Wind Farms in Regions Exposed to Tropical Cyclones, European Wind Energy Association, Milan, Italy. Available online at [https://www.researchgate.net/profile/Samuel_Hernando/publication/228491415_Wind_f arms_in_regions_exposed_to_tropical_cyclones/links/56c6d0d608ae408dfe4e67e0.pdf].

Department of Energy (2015), Wind Vision: A new era for wind power in the United States. Available online at [http://energy.gov/sites/prod/files/2015/03/f20/wv_full_report.pdf].

Fadaeinedjad, R., G. Moschopoulos, and M. Moallem (2009), The Impact of Tower Shadow, Yaw Error, and Wind Shears on Power Quality in a Wind-Diesel System, IEEE Trans. Energy Convers., 24(1), 102–111, doi:10.1109/TEC.2008.2008941.

French, J. R., W. M. Drennan, J. A. Zhang, and P. G. Black (2007), Turbulent Fluxes in the Hurricane Boundary Layer. Part I: Momentum Flux, *J. Atmospheric Sci.*, *64*(4), 1089–1102, doi:10.1175/JAS3887.1.

Hara, T., and P. P. Sullivan (2015), Wave Boundary Layer Turbulence over Surface Waves in a Strongly Forced Condition, *J. Phys. Oceanogr.*, *45*(3), 868–883, doi:10.1175/JPO-D-14-0116.1.

Harper, B. A., J. D. Kepert, and J. D. Ginger (2010), Guidelines For Converting Between Various Wind Averaging Periods in Tropical Cyclone Conditions. Available online at [http://www.wmo.int/pages/prog/www/tcp/documents/WMO_TD_1555_en.pdf].

International Electrotechnical Comission (2007), IEC 61400-1 Wind Turbines- Part 1: Design Requirements, Edition 3, 2007.

Ishihara, T., A. Yamaguchi, and K. Takahara (2005), An Analysis of Damaged Wind Turbines by Typhoon Maemi in 2003., in Proceedings of the sixth Asia-Pacific conference on wind engineering, Seoul, South Korea. Available online at [https://www.researchgate.net/publication/228415873_An_Analysis_of_damaged_wind_t urbines_by_typhoon_Maemi_in_2003].

Jonkman, J. M., and M. L. Buhl Jr. (2005), FAST User's Guide. NREL/EL-500-29798. Golden, CO: National Renewable Energy Laboratory. Available online at [https://nwtc.nrel.gov/FAST]. (Accessed 28 September 2016)

Jordan, M. R., and C. A. Clayson (2008), A new approach to using wind speed for prediction of tropical cyclone generated storm surge, *Geophys. Res. Lett.*, *35*(13), L13802, doi:10.1029/2008GL033564.

Justin Gillis (2016), America's First Offshore Wind Farm May Power Up a New Industry, N. Y. Times, 22nd August, D1. Available online at

[http://www.nytimes.com/2016/08/23/science/americas-first-offshore-wind-farm-may-power-up-a-new-industry.html?_r=0].

Keim, B. D., R. A. Muller, and G. W. Stone (2007), Spatiotemporal Patterns and Return Periods of Tropical Storm and Hurricane Strikes from Texas to Maine, *J. Clim.*, *20*(14), 3498–3509, doi:10.1175/JCLI4187.1.

Kimball, S. K., and M. S. Mulekar (2004), A 15-Year Climatology of North Atlantic Tropical Cyclones. Part I: Size Parameters, *J. Clim.*, *17*(18), 3555–3575, doi:10.1175/1520-0442(2004)017<3555:AYCONA>2.0.CO;2.

Landsea, C. W. (1993), A Climatology of Intense (or Major) Atlantic Hurricanes, *Mon. Weather Rev.*, *121*(6), 1703–1713, doi:10.1175/1520-0493(1993)121<1703:ACOIMA>2.0.CO;2.

National Hurricane Center (2016), Tropical Cyclone Climatology, Available online at [http://www.nhc.noaa.gov/climo/].

Neumann, C., J. (1991), The National Hurricane Center Risk Analysis Program (HURISK), NOAA technical memorandum NWS NHC 38. Available online at [http://www.nhc.noaa.gov/pdf/NWS-NHC-1987-38.pdf].

Park, J., L. Manuel, and S. Basu (2015), Toward Isolation of Salient Features in Stable Boundary Layer Wind Fields that Influence Loads on Wind Turbines, *Energies*, *8*(4), 2977–3012, doi:10.3390/en8042977.

Powell, M. D., and S. Cocke (2012), Hurricane wind fields needed to assess risk to offshore wind farms, *Proc. Natl. Acad. Sci.*, *109*(33), E2192–E2192, doi:10.1073/pnas.1206189109.

Powell, M. D., P. J. Vickery, and T. A. Reinhold (2003), Reduced drag coefficient for high wind speeds in tropical cyclones, *Nature*, *422*(6929), 279–283, doi:10.1038/nature01481.

Richter, D. H., R. Bohac, and D. P. Stern (2016), An Assessment of the Flux Profile Method for Determining Air–Sea Momentum and Enthalpy Fluxes from Dropsonde Data in Tropical Cyclones, *J. Atmospheric Sci.*, *73*(7), 2665–2682, doi:10.1175/JAS-D-15-0331.1.

Russo, G. (2014), Renewable energy: Wind power tests the waters, *Nature*, *513*(7519), 478–480, doi:10.1038/513478a.

Sim, C., S. Basu, and L. Manuel (2012), On Space-Time Resolution of Inflow Representations for Wind Turbine Loads Analysis, *Energies*, *5*(7), 2071–2092, doi:10.3390/en5072071.

Stern, D. P., G. H. Bryan, and S. D. Aberson (2016), Extreme Low-Level Updrafts and Wind Speeds Measured by Dropsondes in Tropical Cyclones, *Mon. Weather Rev.*, *144*(6), 2177–2204, doi:10.1175/MWR-D-15-0313.1.

Suzuki, N., T. Hara, and P. P. Sullivan (2014), Impact of Dominant Breaking Waves on Air–Sea Momentum Exchange and Boundary Layer Turbulence at High Winds, *J. Phys. Oceanogr.*, *44*(4), 1195–1212, doi:10.1175/JPO-D-13-0146.1.

Vanderwende, B. J., and J. K. Lundquist (2012), The modification of wind turbine performance by statistically distinct atmospheric regimes, *Environ. Res. Lett.*, *7*(3), 34035, doi:10.1088/1748-9326/7/3/034035.

Vickery, P. J., and P. F. Skerlj (2005), Hurricane Gust Factors Revisited, *J. Struct. Eng.*, *131*(5), 825–832, doi:10.1061/(ASCE)0733-9445(2005)131:5(825).

Vickery, P. J., D. Wadhera, M. D. Powell, and Y. Chen (2009), A Hurricane Boundary Layer and Wind Field Model for Use in Engineering Applications, *J. Appl. Meteorol. Climatol.*, *48*(2), 381–405, doi:10.1175/2008JAMC1841.1.

Walter, K., C. C. Weiss, A. H. P. Swift, J. Chapman, and N. D. Kelley (2009), Speed and Direction Shear in the Stable Nocturnal Boundary Layer, *J. Sol. Energy Eng.*, *131*(1), 11013, doi:10.1115/1.3035818.

Worsnop, R., G. H. Bryan, J. K. Lundquist, and J. A. Zhang (2016), Spectral and Coherence Characteristics of an LES-modeled Hurricane Boundary Layer for Wind Energy Applications, *Bound.-Layer Meteor.* (in review).

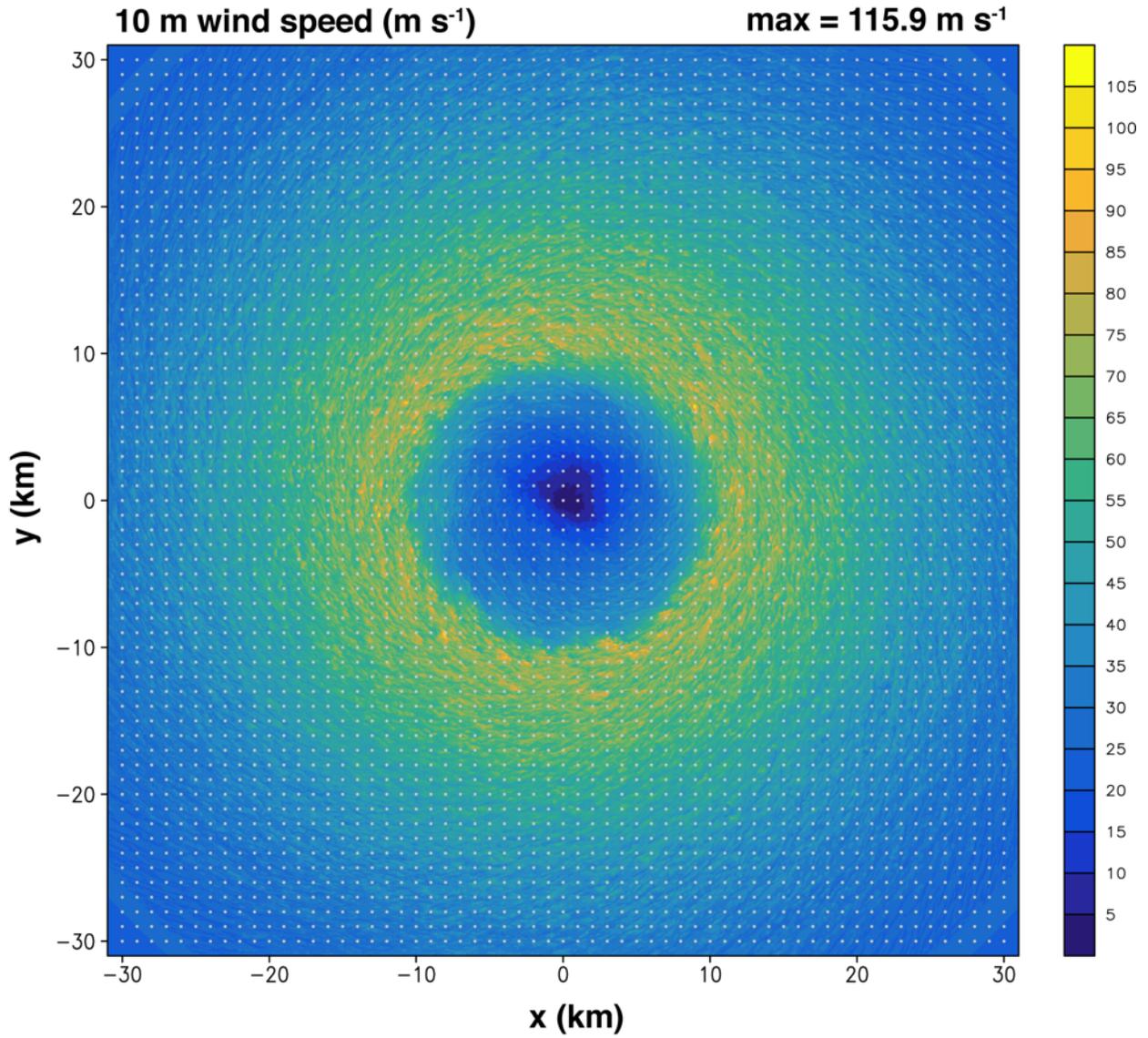

Figure 1 Instantaneous snapshot of the 10-m wind field produced by the CM1 model ($\Delta x = \Delta y = 31.25\ m$. Locations of the mock towers and thus data output are shown as the gray dots.

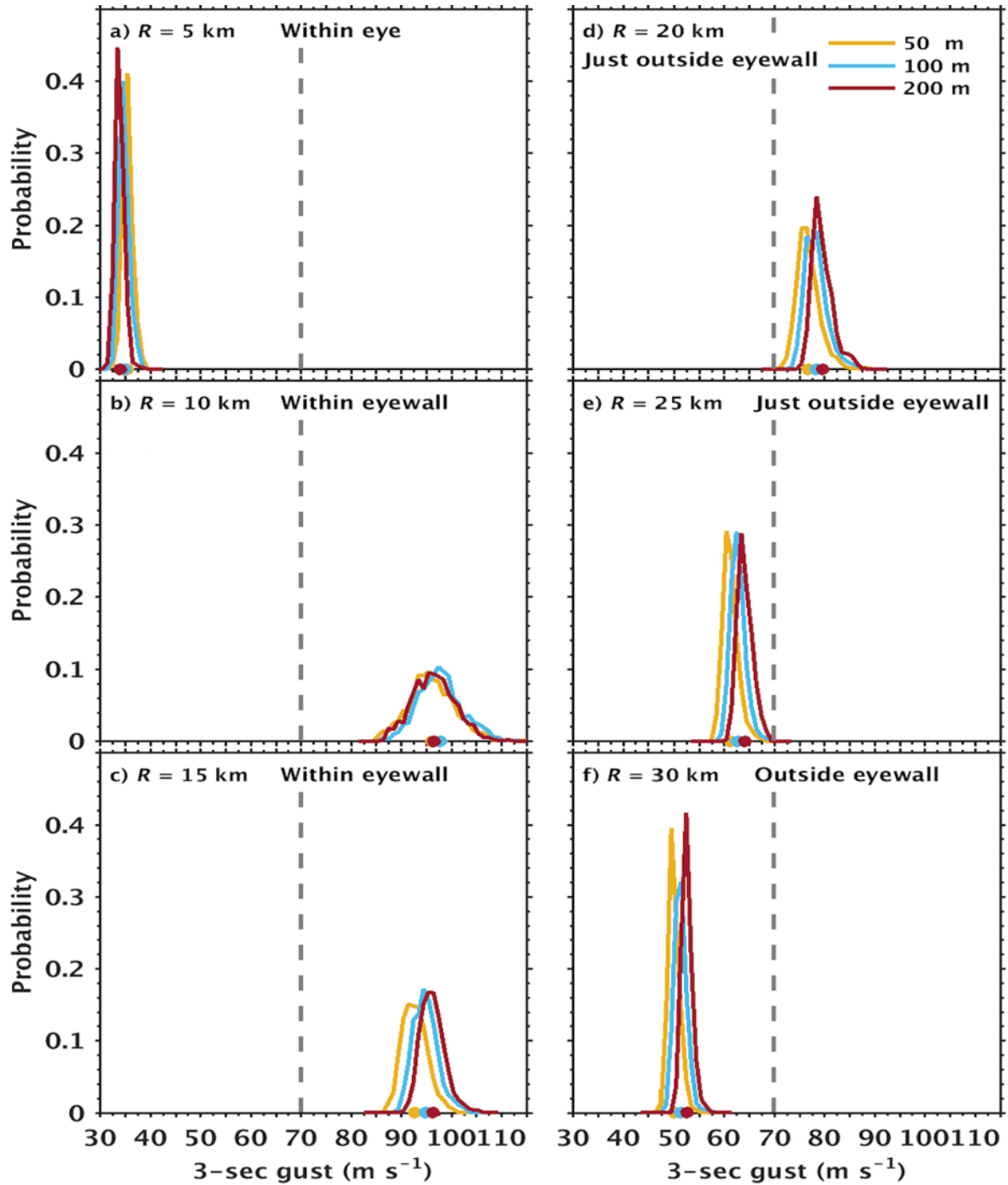

Figure 2 Histograms of the 3-sec gusts at different locations within the hurricane: a) eye (in this case, R = 5 km), b) - c) eyewall (in this case, R = 10 and 15 km), d)-e) just outside of the eyewall (in this case, R = 20 and 25 km, and f) outside of the eyewall (in this case, R = 30 km). Probabilities are shown for gusts at 50 m (gold), 100 m (blue), and 200 m (brown) ASL. Means of the distributions are shown as the gold, blue, and brown dots. For reference, the 70 m s$^{-1}$ gust threshold is also shown (gray dashed).

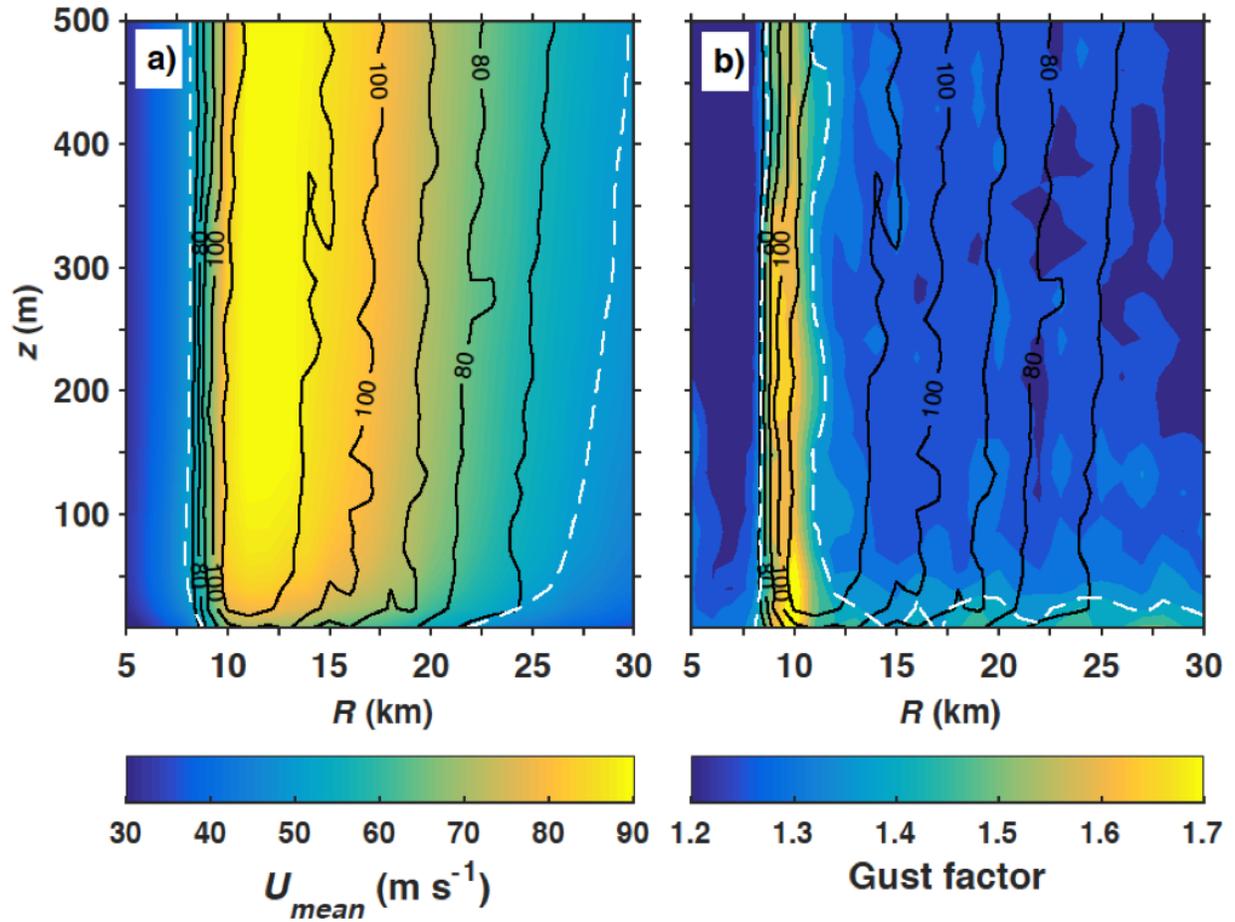

Figure 3 a) Radius-height contours of the maximum 10-min mean wind speed (colored contours) at each radius and height overlaid with maximum 3-sec gusts (black contours, only values exceeding 70 m s$^{-1}$ are plotted). b) Radius-height contours of the maximum gust factor (colored contours) during ten minutes overlaid with maximum 3-sec gusts (black contours, only values exceeding 70 m s$^{-1}$ are plotted). Contours (white-dashed) of the 50 m s$^{-1}$ 10-min mean wind threshold and a threshold gust factor of 1.4 are shown in a) and b), respectively.

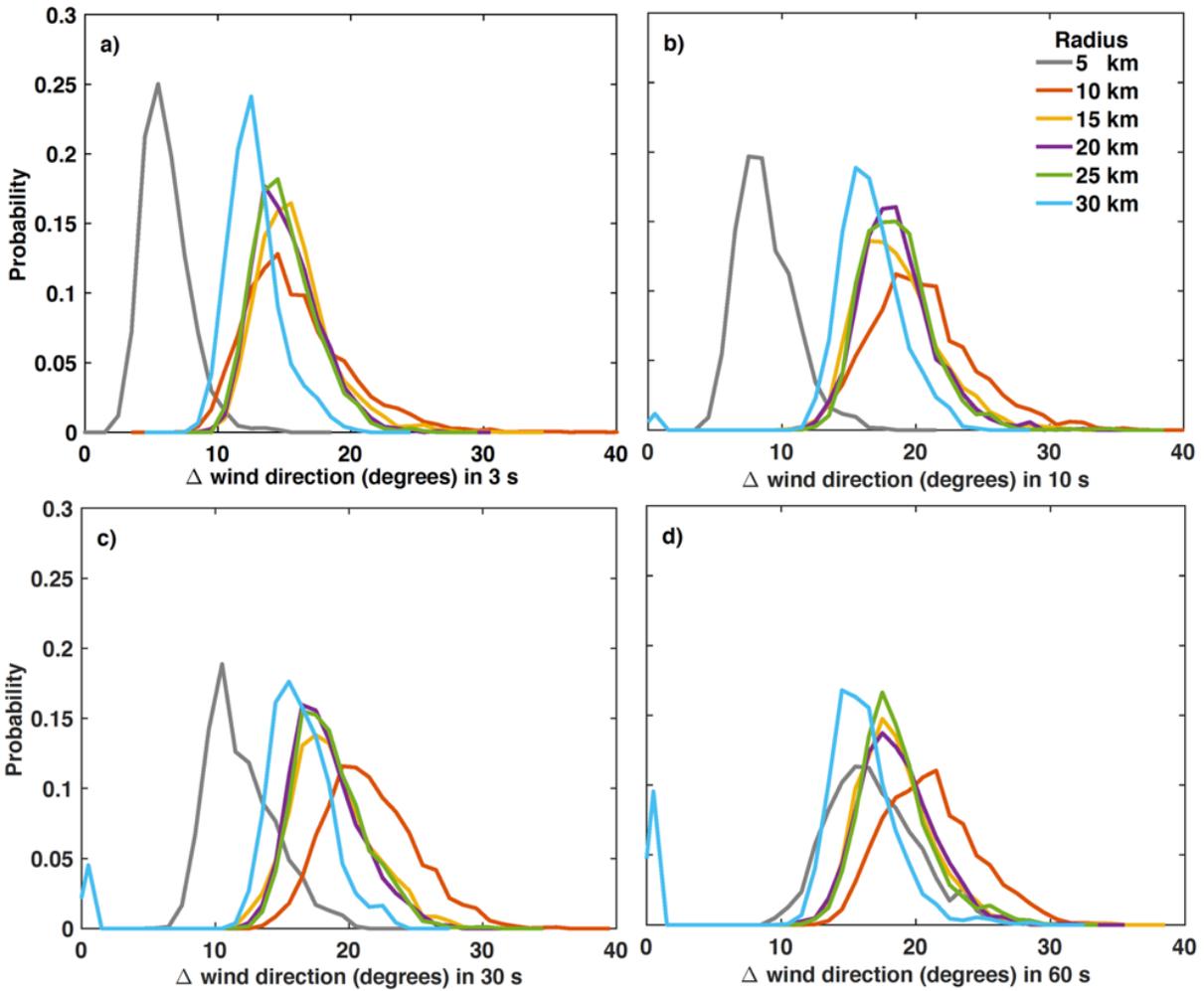

Figure 4 Histograms of the maximum change in wind direction over a) 3, b) 10, c) 30, and d) 60 sec. Six hurricane radii are shown for each histogram: 5 km, 10 km, 15 km, 20km, 25km, and 30 km from the hurricane center.

—

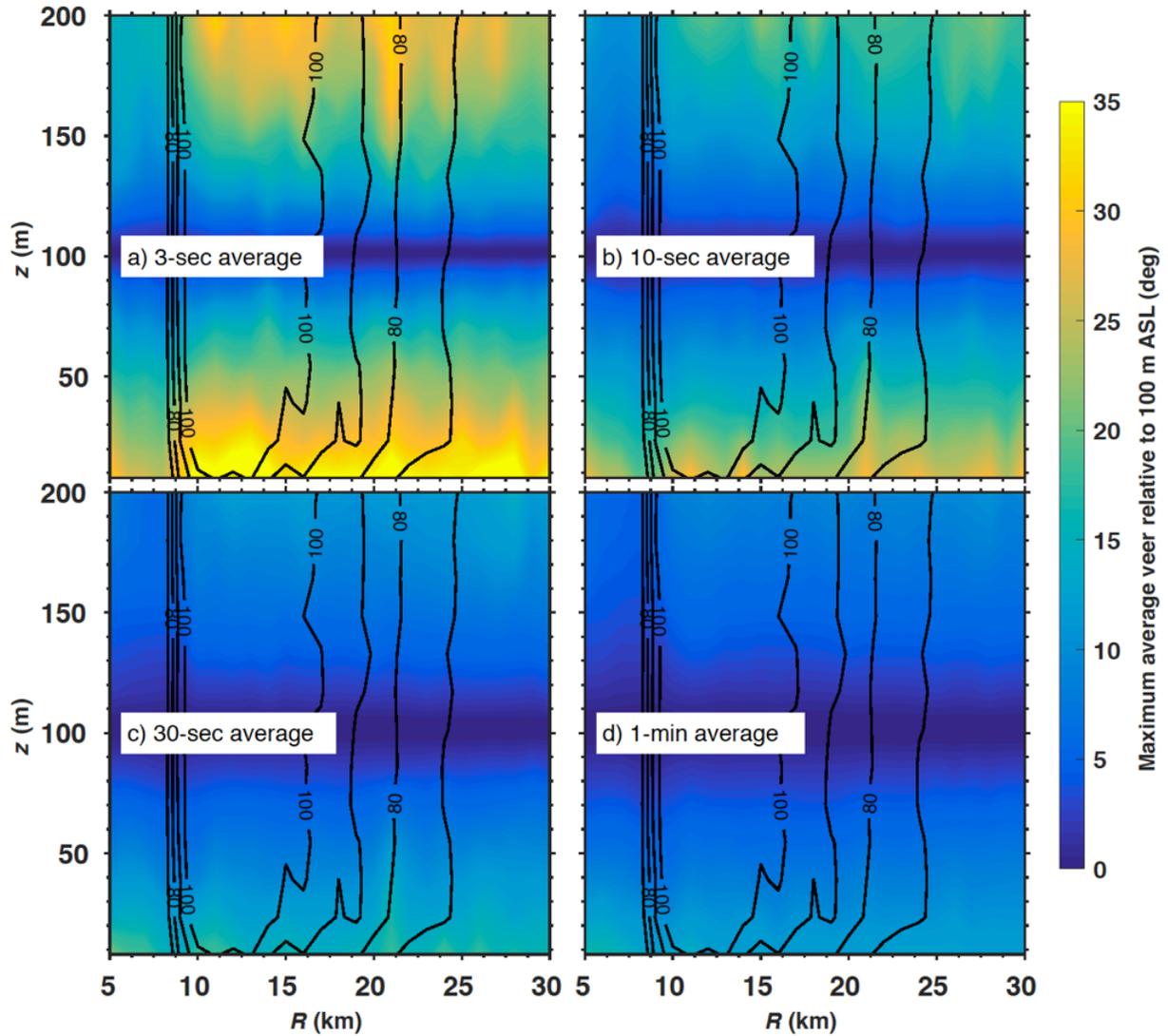

Figure 5 Radius-height contours of the maximum average veer relative to hub height (100 m ASL) (colored contours) for averages calculated over a) 3 sec, b) 10 sec, c) 30 sec, and d) 1 min. Overlaid are the maximum 3-sec gusts (black contours, only values exceeding 70 m s$^{-1}$ are plotted).